# New Comment on Gibbs Density Surface of Fluid Argon: Revised Critical Parameters, L. V. Woodcock, Int. J. Thermophys. (2014) 35, 1770-1784

I. H. Umirzakov

Institute of Thermophysics, Lavrentev prospect, 1, 630090 Novosibirsk, Russia

e-mail: cluster125@gmail.com

*Keywords:* phase equilibrium, first order phase transition, critical point, coexistence, liquid, vapor

## Abstract

It is shown that the existence of single critical point of the fluid described by van der Waals equation of state is not the hypothesis and it is the consequence of the thermodynamic conditions of liquid-vapor phase equilibrium. It is also shown that the thermodynamic conditions of liquid-vapor phase equilibrium of the fluid give equalities to zero of first and second partial derivatives at constant temperature of pressure with respect to volume at critical point which are usual conditions of existence of critical point.

## Introduction

The equilibrium "line of critical states" of liquid-vapor first order phase transition of argon instead of single critical point is thermodynamically defined in very interesting paper [1]. According to [1] the existence of van der Waals critical point [2-5] is based upon a hypothesis rather than a thermodynamic definition and there is no thermodynamic definition of the "critical point" of van der Waals.

In contrast to the conjecture of [1] there is no reliable experimental evidence to doubt the existence of single critical point [6].

In present work we will prove that the existence of single critical point of the fluid described by van der Waals equation of state (VDW-EOS) is not the hypothesis and it is the consequence of the thermodynamic conditions of liquid-vapor phase equilibrium. We also show that the thermodynamic conditions of liquid-vapor phase equilibrium of the fluid give equalities to zero of first and second partial derivatives at constant temperature of pressure with respect to volume at critical point. The equalities are usual conditions of existence of critical point.

## The proof

Let us consider van der Waals equation of state

$$p(T,n) = \frac{nkT}{1-bn} - an^2, \qquad (1)$$

where $a = const > 0$ and $b = const > 0$, $k$ is Boltzmann's constant, $p$ is pressure, $T$ is temperature and $n$ is number density of particles (atoms or molecules) consisting fluid.

The thermodynamic conditions of liquid-vapor phase equilibrium consist of equalities temperatures, pressures and chemical potentials of liquid and vapor coexisting in phase equilibrium. They give for VDW-EOS (1) the following equations [7-11]

$$bkT/a = (1-bn_L)(1-bn_V)(bn_L + bn_V), \qquad (2)$$

$$\ln\left(\frac{1/bn_V - 1}{1/bn_L - 1}\right) = \frac{(bn_L - bn_V)(2 - bn_L - bn_V)}{(1 - bn_L)(1 - bn_V)(bn_L + bn_V)}, \qquad (3)$$

where $n_L = n_L(T)$ and $n_V = n_V(T)$ are number densities of liquid and vapor coexisting in thermodynamic phase equilibrium respectively.

According to [7-11] following relations

$$\frac{bkT(y)}{a} = F(y) \equiv \frac{2(y - e^{2y} + ye^{2y} + 1)(4ye^{2y} - e^{4y} + 1)^2}{(e^{2y} - 1)(2y - 2e^{2y} + e^{4y} - 2ye^{4y} + 4y^2 e^{2y} + 1)^2}, \qquad (4)$$

$$bn_L(T(y)) = F_L(y) \equiv 2 \cdot \frac{y - e^{2y} + ye^{2y} + 1}{(2y-1)e^{2y} - e^{-2y} - 2y + 2}, \qquad (5)$$

$$bn_V(T(y)) = F_V(y) \equiv 2 \cdot \frac{y - e^{2y} + ye^{2y} + 1}{e^{4y} - e^{2y}(2y+2) + 2y + 1}, \qquad (6)$$

where $y$ is the parameter, $0 \le y \le \infty$, are the parametric solution of Eqs. (2) and (3).

Using Eqs. (4)-(6) we obtain from VDW-EOS (1)

$$\frac{b^2 p_e(T(y))}{a} = f(y) \equiv \frac{4e^{2y}(y - e^{2y} + ye^{2y} + 1)^2(e^{4y} - 2e^{2y} - 4y^2 e^{2y} + 1)}{(e^{2y} - 1)^2(2y - 2e^{2y} + e^{4y} - 2ye^{4y} + 4y^2 e^{2y} + 1)^2}, \qquad (7)$$

$$\frac{b^3}{a} \cdot \left.\frac{\partial p(T,n)}{\partial(1/n)}\right|_{n=n_{L,V}(T(y))} = G_{L,V}(y) \equiv \left[-\frac{F(y)}{[1 - F_{L,V}(y)]^2} + 2 \cdot F_{L,V}(y)\right] \cdot F_{L,V}^2(y), \qquad (8)$$

$$\frac{b^3}{a} \cdot \left.\frac{\partial^2 p(T,n)}{\partial(1/n)^2}\right|_{n=n_{L,V}(T(y))} = H_{L,V}(y) \equiv 2 \cdot \left[\frac{F(y)}{[1 - F_{L,V}(y)]^3} - 3 \cdot F_{L,V}(y)\right] \cdot F_{L,V}^3(y), \qquad (9)$$

where $p_e(T) \equiv p(T, n_L(T)) = p(T, n_V(T))$ is the saturation pressure of vapor. Eqs. (7)-(9) corresponds to pressure and its first and second partial derivatives with respect to density at constant temperature along phase equilibrium lines $n_L(T)$ and $n_V(T)$ on the thermodynamic (temperature, density)-plane.

Monotonical functions $F(y)$, $F_L(y)$, $F_V(y)$ and $f(y)$ are shown in Fig. 1 and functions $G_L(y)$, $G_V(y)$, $H_L(y)$ and $H_V(y)$ are shown in Fig. 2.

One can easily see from Figures 1 and 2 or analytically from Eqs. (4)-(9) that there is single thermodynamic state of the fluid corresponding to $y = 0$, where

$$n_L(T(y=0)) = n_V(T(y=0)) = 1/3b, \quad T(y=0) = 8a/27kb, \quad p_e(T(y=0)) = a/27b^2, \qquad (10)$$

$$\left.\frac{\partial p(T,n)}{\partial(1/n)}\right|_{n=n_L(T(y=0))} = \left.\frac{\partial p(T,n)}{\partial(1/n)}\right|_{n=n_V(T(y=0))} = 0, \qquad (11)$$

$$\left.\frac{\partial^2 p(T,n)}{\partial (1/n)^2}\right|_{n=n_L(T(y=0))} = \left.\frac{\partial^2 p(T,n)}{\partial (1/n)^2}\right|_{n=n_V(T(y=0))} = 0. \qquad (12)$$

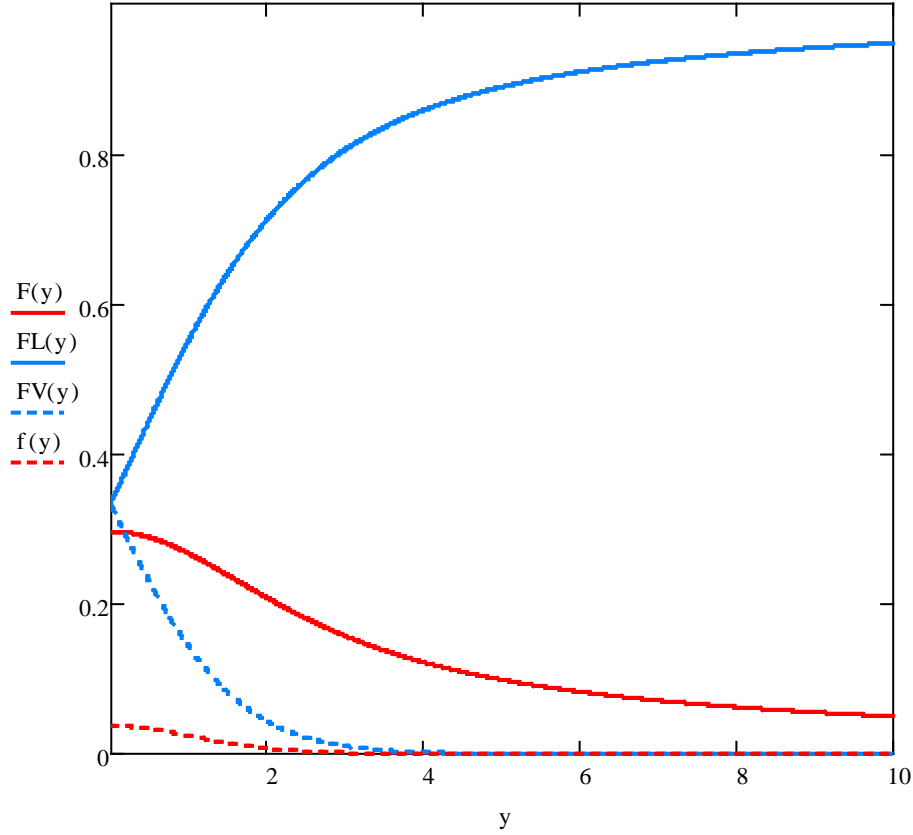

Figure 1. Dependencies of functions $F$, $F_L$, $F_V$ and $f$ on $y$. Solid and dashed red lines correspond to $F$ and $f$ respectively. Solid and dashed blue lines correspond to $F_L$ and $F_V$ respectively.

It is easy to show that the function $p(T,n)$ is analytical function of temperature and number density at $T = 8a/27kb$ and $n = 1/3b$. Therefore the functions $\partial p(T,n)/\partial n$ and $\partial^2 p(T,n)/\partial n^2$ are also analytical functions at $T = 8a/27kb$ and $n = 1/3b$. The value of limit of the analytical function does not depend on the direction of limit [12] therefore

$$\left.\frac{\partial p(T,n)}{\partial (1/n)}\right|_{T=T_c, n=n_c} = 0, \qquad (13)$$

$$\left.\frac{\partial^2 p(T,n)}{\partial (1/n)^2}\right|_{T=T_c, n=n_c} = 0, \qquad (14)$$

where $T_c \equiv 8a/27kb$, $n_c \equiv n_L(T_c) = n_V(T_c) = 1/3b$.

It is easy to see that Eqs. (13) and (14) are exactly the equations which were used in [2] to define the critical point. So $y = 0$ corresponds to the single critical point, where according to Eqs. (10) $T = T_c$, $n = n_c$ and $p = p_c \equiv p_e(T_c) = a/27b^2$.

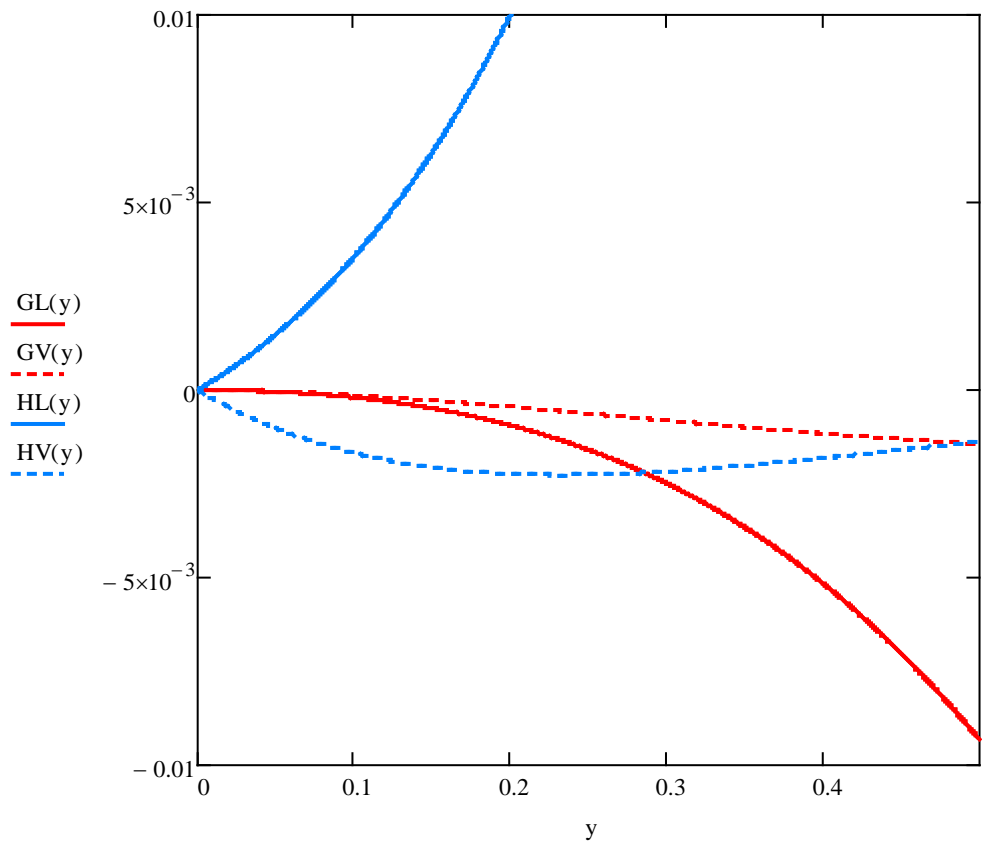

Figure 2. Dependencies of functions $G_L$, $G_V$, $H_L$ and $H_V$ on $y$. Solid and dashed red lines correspond to $G_L$ and $G_V$ respectively. Solid and dashed blue lines correspond to $H_L$ and $H_V$ respectively.

So it is proved that the existence of single critical point of the fluid described by van der Waals equation of state is not the hypothesis and it is the consequence of the thermodynamic conditions of liquid-vapor phase equilibrium. It is also shown that the thermodynamic conditions of liquid-vapor phase equilibrium of the fluid give equalities to zero of first and second partial derivatives at constant temperature of pressure with respect to volume at critical point. The equalities are usual conditions of existence of critical point.